\documentclass[aps,pra,twocolumn,superscriptaddress,showkeys,amsmath,amssymb,longbibliography]{revtex4-1}   
\usepackage{graphicx}
\usepackage{dcolumn}
\usepackage{bm}
\usepackage[colorlinks=true,urlcolor=blue,citecolor=blue,linkcolor=blue]{hyperref}
\expandafter\let\csname equation*\endcsname\relax
\expandafter\let\csname endequation*\endcsname\relax
\usepackage{amsfonts}
\usepackage{lipsum}
\begin{document}

\preprint{APS/123-QED}

\title{A first approach to the open dynamics of bipartite systems}
\thanks{A footnote to the article title}%

\author{M. Salado-Mejía}%
 \email{mariana.salado@academicos.udg.mx}
\affiliation{Departamento de Innovación en Gestión del Conocimiento, CUGDL, Universidad de Guadalajara, Guadalajara, Jalisco, C.P. 44160, México.}%

\date{\today}

\begin{abstract}
In this work, we review the open quantum dynamics of the most known bipartite systems, such as the qubit-qubit system, the oscillator-oscillator system, and the qubit-oscillator system. First, we compare each system with and without rotating wave approximation. In this analysis, we observe the influence of the counter-rotating term in the system dynamics. Also, we compare and analyze the resulting dynamics of the three bipartite systems where, due to the nature of each system, different dynamics are observed, but some similarities are also observed between them. To obtain the system dynamics, we use the same platform, the Qutip Toolbox starting from the phenomenological master equation of each system. We made the latter have the same platform for comparison. We attach the codes to generate these dynamics.
\end{abstract}

\keywords{Qutip,master equation, bipartite systems}
\maketitle


\section{\label{sec:level1}Introduction}

One of the most widely used assumptions in theoretical physics is that the systems are isolated. In reality, the systems are not isolated. All systems interact with other systems and generally interact with an environment. For this reason, it is essential to study open dynamics to understand the influence of the environment on the system.

In quantum physics, we use different tools to study the influences of the environment on the system, as is the Nakajima–Zwanzig equation, the Redfield equation, Caldeira-Leggett model and the Lindblad equation. The form of the last master equation allows us to have a phenomenological form, which can help us to write a master equation without the need to make a rigorous derivation. 

The Lindblad form is known as the most general type of Markovian master equation and satisfies some relevant properties to ensure the physicality of the system, as is the Complete Positive Trace Preserving  (CPTP) \cite{Rivas_2010,Cattaneo_2019}. To describe the system dynamics, the Lindblad master equation uses the reduced density matrix, which describes in a general way the quantum state of a physical system, including the decoherence. As we want to know the dynamics of the system with the influence of the environment, we use the reduced density matrix, which only describes the dynamics of the central system. 


The rest of the paper is structured as follows: In section \ref{sec:level2}, we present the master equations and the Lindblad form in a general way. In Section \ref{sec:level3}, we show and describe the principal bipartite systems.  In Section \ref{sec:level4}, we present the phenomenological master equation of each system, then using the Toolbox Qutip, we show some related results and compare the systems. In section \ref{sec:level5}, the general discussion about the three systems are presented. Finally, we present the conclusions in Section \ref{con}.

\section{\label{sec:level2} Master equation and Lindblad form}

The master equations are tools used to describe the temporal evolution of an open quantum system, this is, the master equation describes the dynamics when the system interacts with an environment. This information is contained in the density matrix. Like the Schrodinger and Von Neuman equations, the master equations are a set of time-dependent differential equations.

In general, the quantum dynamics of the open systems cannot be represented in terms of unitary time evolution operators, as is the case of the Lindblad-type phenomenological master equation.
To clarify, a phenomenological model is a scientific model that consistently describes the fundamental theory but is not directly derived from the theory. \\
We use the Linbland master equation for the density matrix because is known as the most general way to preserve traces and to have a completely positive form of time evolution, this ensures the physical properties of the system. The most general form of the Lindblad equation is as follows:

\begin{align}\label{lindblad}
\dot{\rho}(t)=&-\frac{i}{\hbar}[\hat{H}(t),\rho(t)]\\
&+\sum_n\frac{1}{2}\left[2 C_n \rho(t) C_n^\dagger-\rho(t) C_n^\dagger C_n -C_n^\dagger C_n \rho(t)\right] \nonumber
\end{align}

The first term of the generator represents the unitary part of the dynamics generated by the Hamiltonian $H$. The second part represents the non-unitary part, which shows the dissipation generated by the interaction with the environment \cite{Rivas_2010,Cattaneo_2019}. In this part $C_n=\sqrt{\gamma_n}\hat{A}$ are the collapse operator, where $\hat{A_n}$ are the operators by which the system is coupled with the environment $H_{int}$. To find out how Lindblad's master equation is derived, you can refer to these works.

\section{\label{sec:level3}Systems}

Now, we present the most used bipartite systems and their respective Hamiltonians with and without RWA. These bipartite systems are generated with two basic elements used in quantum mechanics: the qubit and the oscillator. The resulting bipartite systems are the qubit-qubit system, the oscillator-oscillator system, and the qubit-oscillator system. The last one is of special importance because it is used to describe the simplest light-matter interaction system known as the Rabi model. It is fundamental to recognize that each of these systems will have "two" Hamiltonians. The latter is due to the rotating wave approximation (RWA). The rotating wave approximation is used to neglect the terms which oscillate rapidly. These terms also do not conserve the energy on the system. In addition to that, the resulting Hamiltonian is easier to diagonalize.

\subsection{Qubit-Qubit}
\begin{figure}[!htt]
\centering
{\includegraphics[width=5cm, height=2.5cm]{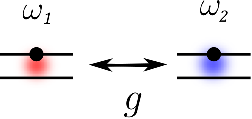}}
\caption{Sketch of qubit-qubit system. Each qubit has its own frequency $\omega_1$ and $\omega_2$, and are coupled with the coupling constant $g$. This system is also used to describe the interaction between two atoms.}\label{atoato}
\end{figure}

The first bipartite system that we will address is the qubit-qubit system. This system is formed by two qubits that interact with each other, as we can see in FIG.\ref{atoato}. This kind of model is used to understand the interaction between two atoms.

The Hamiltonian that describes the whole qubit-qubit interaction is given by:
\begin{equation}\label{qq}
    \hat{H}_{qq}=\frac{\omega_1}{2}\sigma_z^{(1)}+\frac{\omega_2}{2}\sigma_z^{(2)}+g\sigma_x^{(1)} \sigma_x^{(2)}
\end{equation}

\noindent where $\sigma_x^{(i)}$ and $\sigma_z^{(i)}$ are Pauli matrices, the superscripts in the operators indicate the qubit (1 or 2), and $g$ is the coupling constant between the qubits. When the interaction part is developed, we observe four resultant terms. Two terms are co-rotating and, the other two terms are counter-rotating. Using the RWA, we obtain the next Hamiltonian:

\begin{equation}\label{qqrwa}
    \hat{H}_{qq_{RWA}}=\frac{\omega_1}{2}\sigma_z^{(1)}+\frac{\omega_2}{2}\sigma_z^{(2)}+g(\sigma_+^{(1)}\sigma_-^{(2)} +\sigma_-^{(1)}\sigma_+^{(2)}),
\end{equation}

\noindent where $\sigma_{+(-)}$ is the raising (lowering) operator of the qubit. This Hamiltonian has only two interaction terms, both of which conserve the energy of the system.

\subsection{Oscillator-Oscillator}
\begin{figure}[!ht]
\centering
{\includegraphics[width=5cm, height=2.5cm]{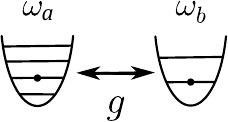}}
\caption{Sketch of oscillator-oscillator system. Each oscillator has its own frequency $\omega_a$ and $\omega_b$, and the oscillators are coupled to each other with the coupling constant $g$.}\label{osciosci}
\end{figure}

Now, we present the oscillator-oscillator system. This system is fundamental in quantum optics because the quantum harmonic oscillator form rises from quantizing the electromagnetic field. Then, the system shown in Fig.\ref{osciosci} can be interpreted as two interacting quantized electromagnetic fields.

To describe the oscillator-oscillator system, we use the following Hamiltonian:
\begin{equation}\label{oo}
\hat{H}_{oo}=\omega_a \hat{a}^\dag \hat{a}+\omega_b \hat{b}^\dag \hat{b}+g(\hat{a}^\dag+\hat{a})(\hat{b}^\dag+\hat{b})
\end{equation}

\noindent  where $\omega_a$ and $\omega_b$ are the oscillator frequencies, $\hat{a}$ ($\hat{b}$) is the annihilation operator of the operator with frequency $\omega_a$($\omega_b$) and, in the similar way that qubit-qubit system, $g$ is the coupling constant between the oscillators.  In this case, the interaction part also generates four terms (co-rotating and counter-rotating terms). Using the RWA, the resultant Hamiltonian is:

\begin{equation}\label{oorwa}
\hat{H}_{oo_{RWA}}=\omega_a \hat{a}^\dag \hat{a}+\omega_b \hat{b}^\dag \hat{b}+g(\hat{a}^\dag\hat{b}+\hat{a}\hat{b}^\dag)
\end{equation}

\noindent where this Hamiltonian only maintains the co-rotate terms that conserve the energy.

\subsection{Oscillator-Qubit}

\begin{figure}[!ht]
\centering
{\includegraphics[width=4.5cm, height=2.2cm]{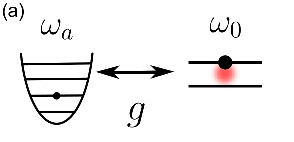}}
{\includegraphics[width=4cm, height=2.5cm]{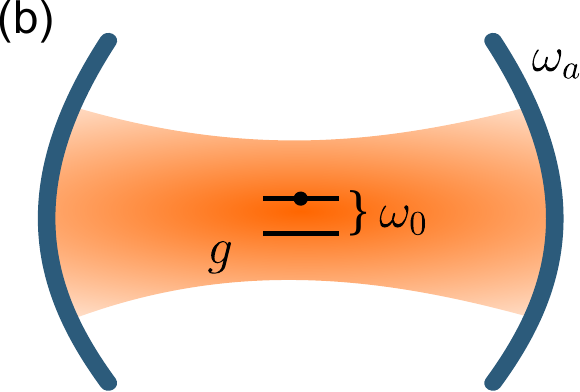}}
\caption{(a) Sketch of the qubit-oscillator system. (b) Sketch of a cavity interacting with a single two-level atom. The cavity and the atom have different frequencies and, they are coupled with the coupling constant $g$.}\label{rabi}
\end{figure}

So far, the bipartite systems presented are composed of the same elements, i.e., only qubits or only oscillators. Now, we show a mixed system: the qubit-oscillator system. The latter system is used to model the interaction between the quantized electromagnetic field and a two-level atom being the most basic light-matter interaction system, as we can observe in FIG.\ref{rabi}. This model is known as the Rabi model \cite{Braak2011} and, it is a fundamental model in quantum optics.

The Hamiltonian that describes the whole qubit-oscillator interaction is given by:
\begin{equation}\label{qo}
    \hat{H}_{qo}=\frac{\omega_0}{2}\sigma_z+\omega_a\hat{a}^\dag\hat{a}+g(\hat{a}+\hat{a}^\dag)\sigma_x
\end{equation}

\noindent where $\omega_0$ is the qubit frequency (two-level atom frequency), $\omega_a$ is the oscillator frequency (cavity frequency), $\sigma_x$, $\sigma_z$, $\hat{a}$ are operators which have been described before.  As in the other systems, $g$ is the coupling constant. Using the RWA, the resulting Hamiltonian is:

\begin{equation}\label{qorwa}
\hat{H}_{qo_{RWA}}=\frac{\omega_0}{2}\sigma_z+\omega_a\hat{a}^\dag\hat{a}+g(\hat{a}\sigma_+ +\hat{a}^\dag \sigma_{-}).
\end{equation}

This Hamiltonian is well-known as the Jaynes-Cummings Hamiltonian. Unlike the Hamiltonian \eqref{qo}, this Hamiltonian is easier to diagonalize.

\section{\label{sec:level4}Master equation for each system and results}

In this section, we present the Lindblad phenomenological master equation for each system with its respective dynamics. To find the dynamics of the systems, we solve numerically the master equations using the Qutip Toolbox in Python.

\subsection{Qubit-Qubit}
Using the Eq.\eqref{lindblad} and defining the collapse operators, we can write the phenomenological Lindblad master equation for the qubit-qubit system. In this case the operators which the system are coupled with the environment are $\sigma_+^{(1)}$,$\sigma_-^{(1)}$,$\sigma_+^{(2)}$ and $\sigma_-^{(2)}$. Then, the master equation can be written as:

\begin{align}\label{meqq}
\dot{\rho}=-i[\hat{H}_{j}, \rho]&+\gamma (N_\omega L[\sigma_+^{(1)}]+ (N_\omega+1)  L[\sigma_-^{(1)}]) \\
& + \kappa(N_\omega  L[\sigma_+^{(2)}]+(N_\omega+1) L[\sigma_-^{(2)}]), \nonumber
\end{align}

\noindent where the sub index $j=qq, qq_{RWA}$ are used to distinguish the Hamiltonian without RWA (Eq.\eqref{qq}) and with RWA(Eq.\eqref{qqrwa}), $N_\omega$ is the temperature in frequency units\cite{Carmichael1993,BreuerHeinz-Peter2002}, $\kappa$ is the dissipation rate of qubit $1$ and $\gamma$ is the dissipation rate of qubit $2$. 

Having the master equation, we can employ the Toolbox Qutip to solve this differential equation. To use the function $mesolve()$ of Qutip, we need the Hamiltonians Eq.\eqref{qq} and Eq.\eqref{qqrwa} including their parameters, the initial state, a time list, the list of collapse operators given in the master equation Eq.\eqref{meqq} and finally, the operators for the expectation values.

\begin{figure}[ht!]
\centering
{\includegraphics[width=9cm, height=5.5cm]{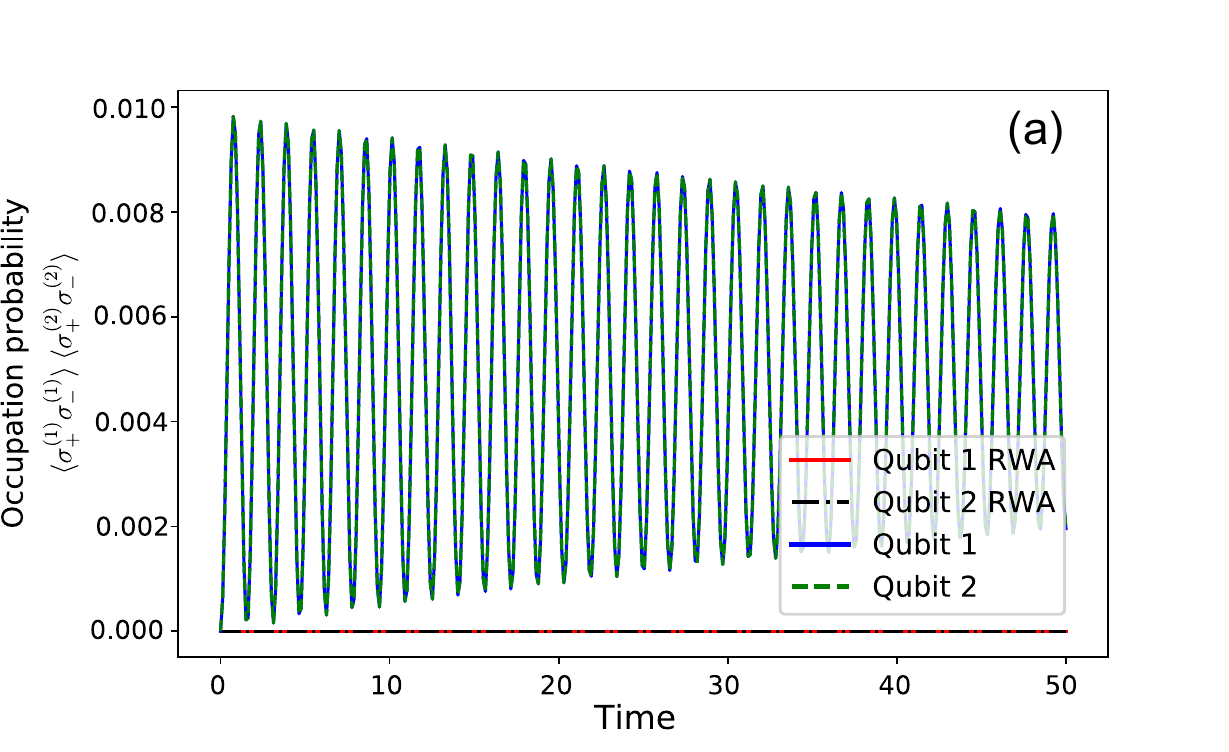}}
{\includegraphics[width=9cm, height=5.5cm]{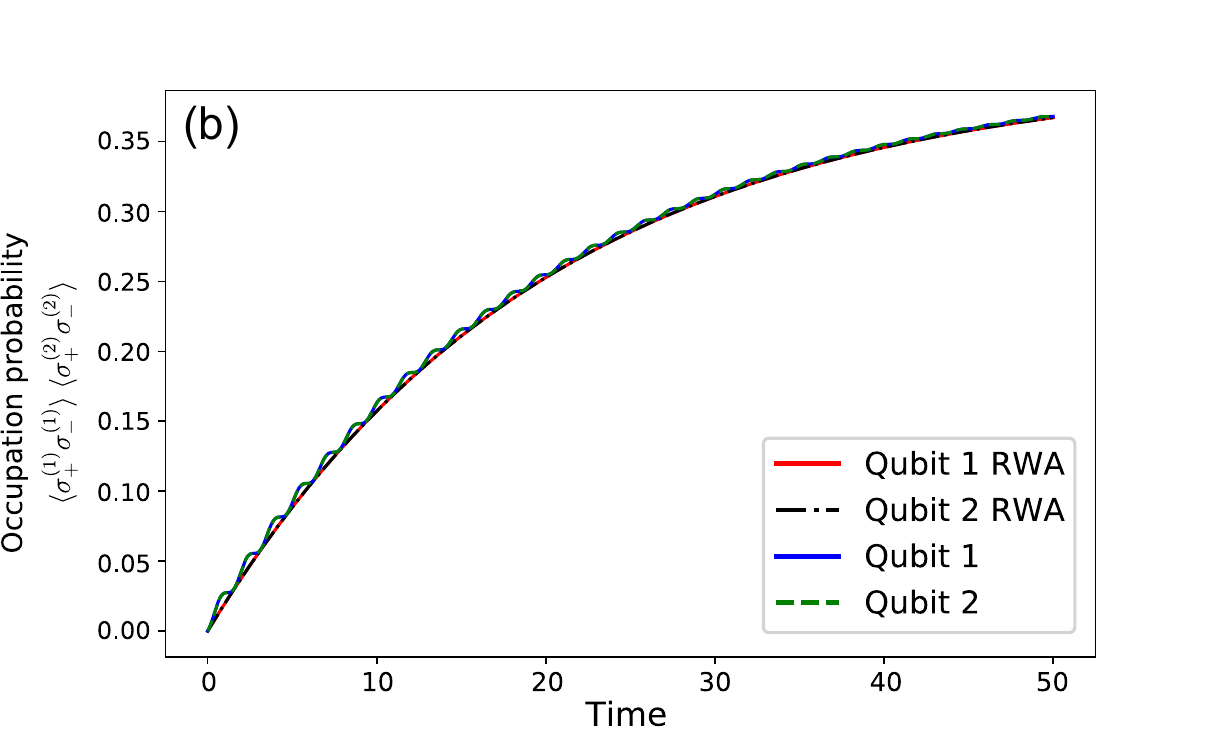}}
\caption{Occupation probability of qubit-qubit system $\langle \sigma_+^{(i)} \sigma_-^{(i)}\rangle$. (a) System at $T=0$. (b) System at $T=2$. The frequencies are at resonance $\omega_1=\omega_2=1$, and the coupling constant is $g=0.2$. The dissipation rates are $\gamma=\kappa=0.01$ and the initial state is $|0 0 \rangle$. Using Eq.\eqref{qqrwa}, the red solid line and black dot-dashed
line correspond to qubit 1 and qubit 2, respectively. Using  Eq.\eqref{qq}, the blue solid line and green dashed line correspond to qubit 1 and qubit 2, respectively. }\label{soqq}
\end{figure}

Then, the first solution is presented in FIG \ref{soqq} (a), where is shown the occupation probability or the value expectation for each qubit using the average $\langle \sigma_+^{(i)} \sigma_-^{(i)}\rangle$ at $T=0$. The latter temperature value simplifies the master equation to:

\begin{equation}
\dot{\rho}=-i[\hat{H}_{j}, \rho]+\gamma  L[\sigma_-^{(1)}] + \kappa L[\sigma_-^{(2)}], \label{meqqt0}
\end{equation}\noindent

This equation is relevant because although the environment temperature is zero, the system still exhibits dissipation, as we can also observe in Fig. \ref{soqq}(a). In this figure, we also notice the differences between the system with RWA and the system without RWA. In the solution generated by the Hamiltonian Eq. \eqref{qq}, no oscillation is observed, and the occupation probability of both qubits is maintained at 0, which implies that the system does not have any dynamics. On the other hand, in the solution generated by the Hamiltonian Eq. \eqref{qqrwa}, we notice that both qubits show dynamics despite being both in the ground state, which is due to the counter-rotating terms. Another considerable behavior is the presence of oscillation in the dynamics. These oscillations rise and depend on the coupling constant $g$, as you can see in \cite{Joshi2014} . Due to the dissipation rate, we observe a decrease in the occupation probability until to reach the steady state.

To study the system when $T>0$ as we can see in the FIG\ref{soqq}(b), we employ the same parameters as the used in Fig.\ref{soqq}(a), but the environment has a temperature $T=2$. In this figure, we can notice the influence of the temperature in the system, and we observe a rapid increase in the probability of occupation at the beginning. The system reaches the steady-state approximately at $\langle \sigma_+^{(i)} \sigma_-^{(i)}\rangle=0.45$. In this figure, we also observe an oscillation in the occupation probability when we use the full Hamiltonian (Eq. \eqref{qq}).

Comparing FIG.\ref{soqq}(a) with FIG.\ref{soqq}(b), we observe that the system with $T\neq 0$ tends more quickly to the steady-state. In both figures, we observe oscillations when the whole interaction is used (Eq. \eqref{qq}), but in  FIG.\ref{soqq}(b), these oscillations vanish faster. Then, in both figures, we observe the influence of the environment and the counter-rotating terms in the system dynamics.

\subsection{Oscillator-Oscillator}
In the same way that we write the master equation for the qubit-qubit system, we can write the phenomenological Lindblad master equation for the oscillator-oscillator system, and it is given by:

\begin{align}
\dot{\rho}=-i[\hat{H}_{k}, \rho]&+\kappa (N_\omega  L[\hat{a}^\dagger]+ (N_\omega+1) L[\hat{a}])\\
&+ \gamma (N_\omega   L[\hat{b}^\dagger]+(N_\omega+1) L[\hat{b}])\nonumber, \label{meoo}
\end{align}

\noindent where the sub index $k=oo, oo_{RWA}$ is used to distinguish the Hamiltonian without (Eq.\eqref{oo}) and with RWA(Eq.\eqref{oorwa}), $N_\omega$ is mean thermal occupation number, $\kappa$ is the dissipation rate of oscillator $a$ and $\gamma$ is the dissipation rate of oscillator $b$.

\begin{figure}[!ht]
\centering
{\includegraphics[width=9cm, height=5.5cm]{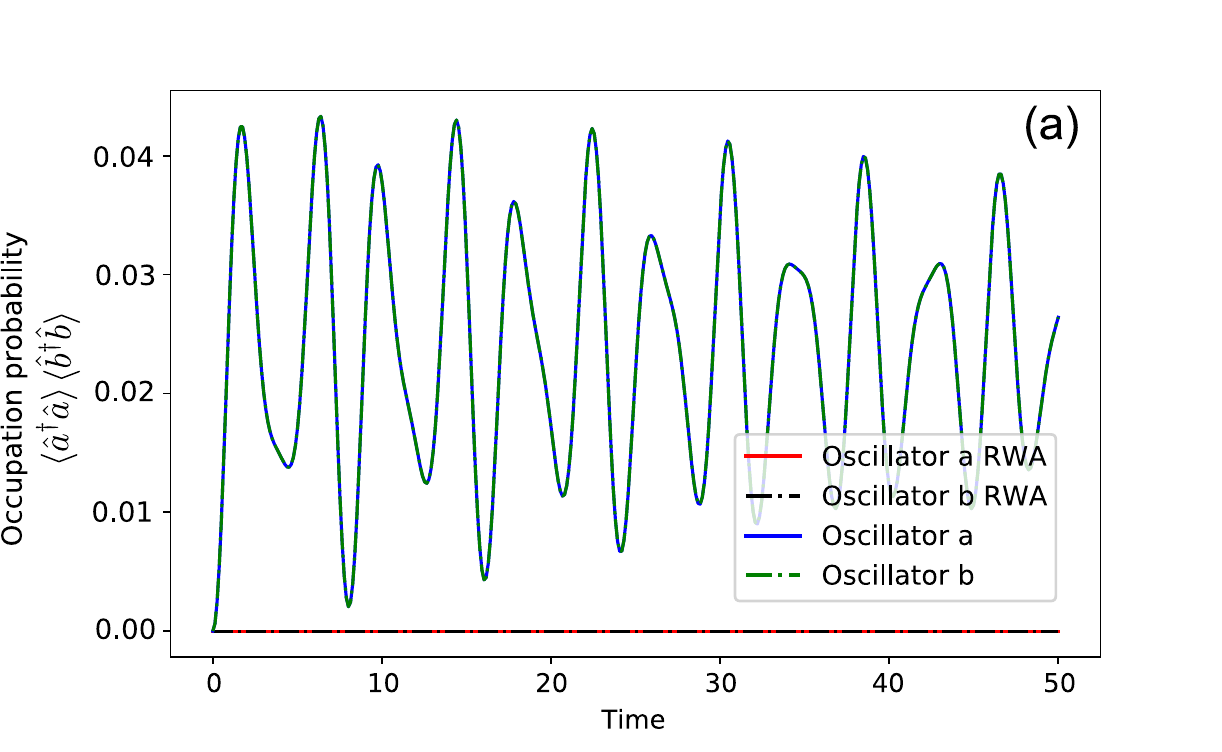}}
{\includegraphics[width=9cm, height=5.5cm]{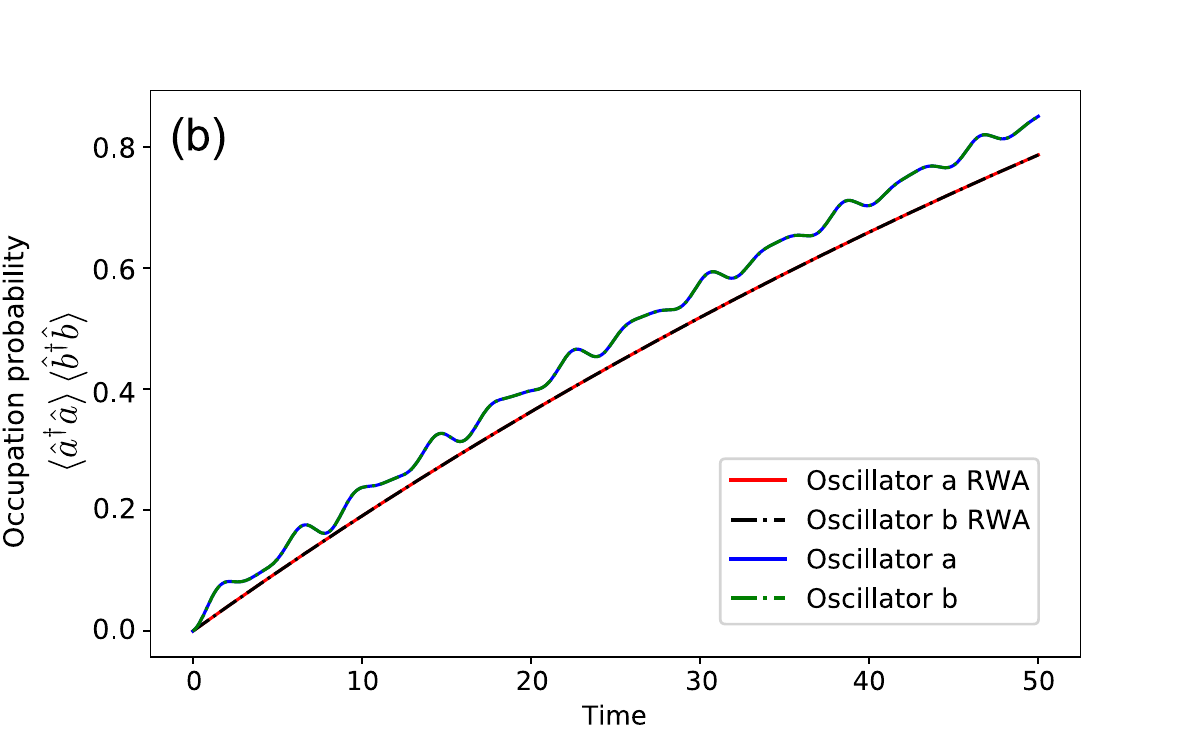}}
\caption{Occupation probability of oscillator-oscillator system using the average excitation numbers $\langle \hat{a}^\dagger\hat{a}\rangle$ and  $\langle \hat{b}^\dagger\hat{b}\rangle$. (a) System at $T=0$. (b) System at $T=2$. The frequencies are at resonance $\omega_a=\omega_b=1$, and the coupling constant, the dissipation rates, and the initial state are the same employed in FIG.\ref{soqq}.  The red solid line and black dot-dashed line correspond to oscillator $a$ and oscillator $b$, respectively, using Eq.\eqref{oorwa}. Blue solid line and green dashed line correspond to oscillator $a$ and oscillator $b$, respectively, using Eq.\eqref{oo}.}\label{sooo}
\end{figure}

Likely to the procedure for the qubit-qubit system, we get the dynamic for the oscillator-oscillator system. When the temperature of the environment is $T=0$, as is shown in FIG \ref{sooo}(a). In this figure, we observe the occupation probability for each oscillator using the average $\langle \hat{a}^\dagger\hat{a}\rangle$ and  $\langle \hat{b}^\dagger\hat{b}\rangle$ at $T=0$. Also, when the temperature is $T=0$, the master equation is simplified:

\begin{equation}
\dot{\rho}=-i[\hat{H}_{k}, \rho]+\kappa L[\hat{a}]+ \gamma L[\hat{b}], 
\end{equation}

\noindent which implies the existence of the dynamics although there is no temperature.

In FIG \ref{sooo}(a), we also observe a difference between the system with RWA and the system without RWA.
The system without RWA shows an oscillation in their dynamics while the system with RWA has no dynamics, similar to the qubit-qubit system using the RWA at $T=0$, as is perceived in FIG \ref{soqq}(a). In the following subsection, we discuss it.  However, in FIG \ref{sooo}(a), the occupation probability exhibits a continuous increase with small oscillations superimposed on the growth. Compared with FIG \ref{sooo}(a), where well-defined oscillations damp out and the system reaches a steady-state at $\langle \hat{a}^\dagger\hat{a}\rangle=\langle \hat{b}^\dagger\hat{b}\rangle \approx 0.025$, in FIG \ref{sooo}(b) the oscillations persist throughout the simulated time but with much smaller relative amplitude due to the dominant thermal growth. The system does not reach a steady-state within the simulated time frame ($t=0-50$), instead reaching occupation values of approximately 0.85 at t=50. This behavior indicates that at finite temperature ($T=2$), the continuous energy input from the thermal environment drives the system away from equilibrium on the simulated timescale. The latter confirms that temperature significantly influences both the dynamics and the damping characteristics of the system."

Finally, comparing FIG \ref{sooo}(a) with FIG \ref{sooo}(b), we observe the influence of the temperature in the environment. Specifically, we will pay attention to the dynamics generated by counter-rotating terms. In FIG \ref{sooo}(a), the occupation probability oscillates and dampens until it reaches the steady-state at $\langle \hat{a}^\dagger\hat{a}\rangle=\langle \hat{b}^\dagger\hat{b}\rangle \approx 0.025$. However, in FIG \ref{sooo}(b), the occupation probability has an increasing and oscillations in their dynamics, but these oscillations disappear faster compared with FIG \ref{sooo}(b). The latter indicates that the temperature influences the damped rate.

\subsection{Qubit-Oscillator}
Using the Eq\eqref{lindblad} and defining the collapse operators, we write and present the phenomenological master equation for the qubit-oscillator system, which is given by:
\begin{align}\label{meqo}
\dot{\rho}=-i[\hat{H}_{l}, \rho]&+\kappa(N_\omega L[\hat{a}^\dagger]+ (N_\omega+1)L[\hat{a}])\\
&+\gamma(N_\omega L[\sigma_+]+(N_\omega+1) L[\sigma_-]),\nonumber 
\end{align}

\noindent where the sub-index $l=qo, qo_{RWA}$  is used to distinguish the Hamiltonian without RWA (Eq.\eqref{qo}) and the Hamiltonian with RWA(Eq.\eqref{qorwa}), the $N_\omega$ is the temperature in frequency units, $\kappa$  is the dissipation rate of the oscillator (cavity) and, $\gamma$ is the dissipation rate of the qubit (an atom with two-levels).

\begin{figure}[!ht]
\centering
{\includegraphics[width=9cm, height=5.5cm]{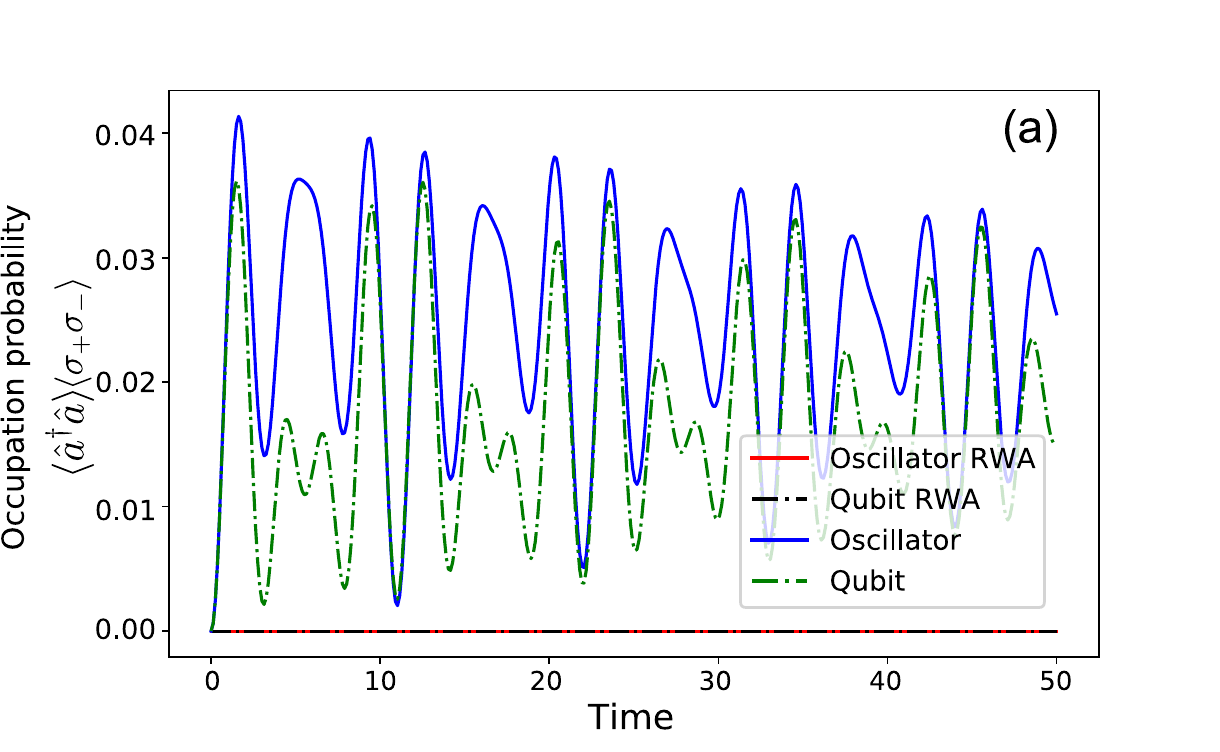}}
{\includegraphics[width=9cm, height=5.5cm]{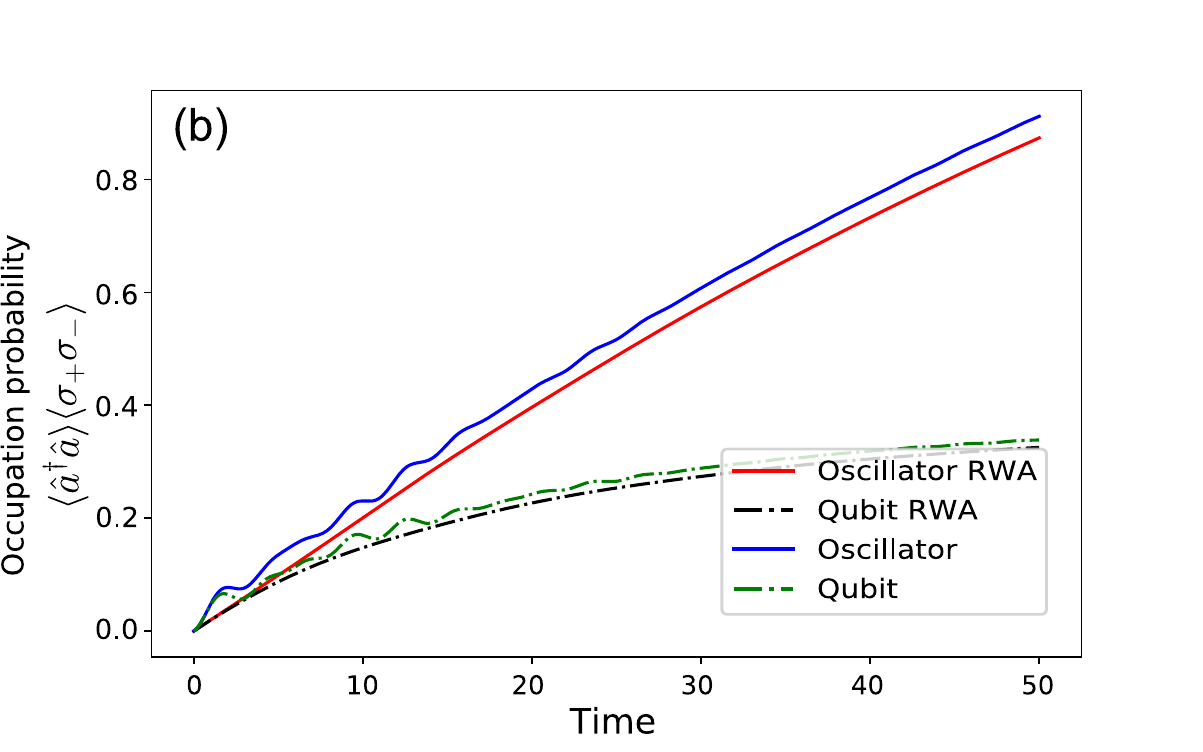}}
\caption{Occupation probability of qubit-oscillator system using the average excitation numbers $\langle \hat{a}^\dagger\hat{a}\rangle$ and  $\langle \sigma_+ \sigma_- \rangle$. (a) System at $T=0$. (b) System at $T=2$. The frequencies are at resonance $\omega_a=\omega_0=1$. The coupling constant, the dissipation rates, and the initial state are the same as in FIG.\ref{soqq}.  The red solid line and black dot-dashed line correspond to the oscillator (cavity) and the qubit (atom), respectively, using Eq.\eqref{qorwa}. Blue solid line and green dashed line correspond to oscillator and qubit, respectively, using Eq.\eqref{qo}.}\label{soqo}
\end{figure}

In FIG \ref{soqo}, we show the occupation probability for the oscillator and the qubit using the average excitation numbers $\langle \hat{a}^\dagger\hat{a}\rangle$ and $\langle \sigma_+ \sigma_-\rangle$, respectively.  The FIG \ref{soqo}(a) shows the occupation probability at $T=0$. When the system only uses the rotating terms (Eq.\eqref{qorwa}), we don't observe any dynamics in the system, in fact, $\langle \hat{a}^\dagger\hat{a}\rangle=\langle \sigma_+ \sigma_-\rangle=0$ through the time. However, when the system has rotating and counter-rotating terms (Eq.\eqref{qo}), we notice astonishing behavior. The dynamics of the oscillator(cavity) presents oscillations, and these oscillations show a decrease as time passes, but this decrease is not monotonous. Sometimes the oscillations are higher or sometimes lower than the previous oscillation. Now, comparing the qubit dynamic with the oscillator dynamics, we note that the dynamics are too different, contrary to the other systems. In the qubit dynamics, we observe a decreasing oscillations, which seems the sum of two oscillations with different frequencies, even though both system elements are in resonance. The latter is due to the nature of each system element, one of them is a system with infinite energy levels(oscillator), and the other has only two energy levels(qubit). Rewriting the master equation Eq.\eqref{meqo} at $T=0$ we get:

\begin{equation}
\dot{\rho}=-i[\hat{H}_{l}, \rho]+\kappa L[\hat{a}]+\gamma L[\sigma_-].
\end{equation}

Now, we analyze the FIG. \ref{soqo}(b). In this figure, we present the occupation probability for the qubit-oscillator system at $T=2$. 
When the system uses the RWA, we observe that the occupation probability for each part of the system does not present a relevant dynamic. Only it is noted the dynamics given by the environment. When we work without the RWA, we observe some small oscillations in each part of the system. These oscillations damp out very quickly, indicating a direct influence from the environment.

Comparing both figures, we observe the influence of temperature in the system dynamic. First, the oscillations in FIG \ref{soqo}(a) are dampened slowly in comparison with the FIG \ref{soqo}(b). In the qubit dynamics, we also observe that the steady-state in FIG \ref{soqo}(b) is reached very quickly ($t\approx 10$), while FIG \ref{soqo}(b), although damping is observed, the steady-state is not observed within the times that are handled. Then, we can conclude that the temperature quickly damps the oscillations of the system dynamic.

\section{General discussion}
Here, we compare the open dynamics of the three presented systems. For simplicity, in all master equations, we define both dissipation values as the same ($\kappa=\gamma$), the systems are at resonance ($\omega_1=\omega_2$, $\omega_a=\omega_b$ and $\omega_0=\omega_a$), and the initial state is set as the ground state ($|0 0\rangle$). 

First, we compare the system dynamics at $T=0$, which means compare FIG \ref{soqq}(a), FIG \ref{sooo}(a) and FIG \ref{soqo}(a).  Observing the occupation probabilities that only use the rotating term, we notice that no system presents dynamics. This behavior is expected because the initial state is the ground state for each element. When the counter-rotating terms are present, the dynamics of the three systems present some similarities, such as they present oscillations, and as time passes by, the oscillations are dampened until reaching the steady-state. Specifically, the figures FIG \ref{soqq}(a), FIG \ref{sooo}(a) exhibit more similarities.  Comparing the nature of the elements for each system, we notice that they are the same for each system, i.e., we have two qubits or two oscillators, and both elements are in the ground state. The latter generated identical dynamics for each element. While in the FIG \ref{soqo}(a) presents the dynamics for a mixed system (a qubit and an oscillator), and each element presents different dynamics. The main difference observed between the three systems is that each one of them shows a different oscillation frequency and a different damping speed, even though they all have the same frequency in their elements, the same dissipation ratio, and the same coupling constant. Observing the qubit-oscillator system, we notice that it is the system with the highest frequency and the highest damping. And the qubit-qubit system is the one with the lowest frequency and the lowest damping. We attribute these differences to the nature of each system.

Now,  we compare the system dynamics at $T=2$, which means we compare FIG \ref{soqq}(b), FIG \ref{sooo}(b) and FIG \ref{soqo}(b). In these three figures, we observe the influence of the temperature in the dynamics, and it produces some similarities in the occupation probabilities. The most relevant similarity observed is the exponential growth (type $ 1-a e ^ {- b x} $), but each system presents an increase with different slopes until it reaches the steady-state. As in figures (a)'s, we observe that systems with the same nature exhibit the same dynamics. We note that the most striking dynamics are generated by the systems with counter-rotating terms since these dynamics also present oscillations. But these oscillations are dampened faster compared with the oscillations on the figures (a)'s. Then, we can conclude that the temperature in the environment damps faster the oscillations on the systems, and also produces more excitations as is shown in the exponential growth of the occupation probabilities.

An interesting question is, Is it valid to apply for the usual average excitation number when the counter-rotating terms are used? c Some works mentioned that the application of the standard procedure to the ultrastrong coupling regime would predict unphysical excitations for a system in its ground state which contains a finite number of excitations due to the counter-rotating terms in the Hamiltonian \cite{FriskKockum2019, Beaudoin2011}, as we observe in figures FIG \ref{soqq}(a), FIG \ref{sooo}(a) and FIG \ref{soqo}(a). 
Despite the above, we found some similar behavior with this works, such as the oscillation due to the counter-rotating terms. It is important to remark that our work presents a first approximation of the open dynamics of the most known bipartite system and their respective comparison under the same platform.

We note that the phenomenological local master equations used here are valid in the weak system-environment coupling regime . In the strong coupling regime, local master equations may fail to capture dissipative critical behavior or violate thermodynamic consistency . More rigorous derivations, such as global master equations or reaction-coordinate mappings, are required in that regime \cite{Konopik2020, trushechkin2021quantum, Hofer_2017}. Our choice is justified by the weak dissipation rates ($\gamma, \kappa \ll \omega$) employed in this study.

It is important to note that the validity of standard observables such as $\langle \hat{a}^\dagger\hat{a}\rangle$ and $\langle \sigma_+ \sigma_-\rangle$ is limited to the regime $g/\omega \leq 0.3$ . In the ultrastrong coupling regime, these observables may predict unphysical excitations due to the ground state containing virtual photons . In our simulations, we set $g/\omega = 0.2$, which lies within the valid regime for standard observables \cite{FriskKockum2019}.

\section{\label{con}Conclusions}

In this work, we have shown the open dynamics of the main bipartite systems using the phenomenological master equation. To solve the phenomenological master equations, we use a numerical solver from the Toolbox Qutip on Python. Using the phenomenological master equations and the same numerical solver, we created a common platform that helps us compare the dynamics to be valid, finding similarities and differences between the systems. The most relevant similarity that we found is when systems have elements of the same nature, they present similar dynamics, such as the dynamics will present damped oscillations until reaching the steady-state. On the other hand, the differences that we found are that the oscillation frequency and the damping speed are related to the nature of the system. The most striking dynamics are found in the mixed system (qubit-oscillator), in which each element exhibits different dynamics. Furthermore, this system shows the widest oscillation frequency and the fastest damping speed. As a first approximation, we present the influence of the counter-rotating terms in the dynamics system, which produces the oscillations on the dynamics.

In addition to the fact that the mixed system (qubit-oscillator) is the one that presents the most interesting dynamics in each of its elements since they are different and is the one with the widest oscillation frequency and the fastest damping speed. Furthermore, we show that the study of the open dynamics of these systems serves as a first approximation to know the importance of co-rotating and counter-rotating terms in the system using a common tool.

\appendix

\nocite{*}

\bibliography{nref}

\end{document}